\documentclass[aps,pre,twocolumn,longbibliography]{revtex4-1}

\usepackage{amsmath,amssymb}
\usepackage{mathrsfs}
\usepackage{graphicx}
\usepackage{xcolor}
\usepackage{dsfont}
\usepackage[hidelinks]{hyperref}
\usepackage[normalem]{ulem}
\usepackage[utf8]{inputenc}

\hypersetup{colorlinks=true, linkcolor=black, citecolor=black, urlcolor=blue}

\begin{document}

\title{Inverted harmonic oscillator dynamics of the non-equilibrium phase transition in the Dicke model}
\author{Karol Gietka}
\email[Corresponding author: ]{karol.gietka@oist.jp}
\author{Thomas Busch}
\affiliation{Quantum Systems Unit, Okinawa Institute of Science and Technology Graduate University, Onna, Okinawa 904-0495, Japan} 
\date{\today}

\begin{abstract}
We show how the dynamics of the Dicke model after a quench {from the ground-state configuration of the normal phase} into the superradiant phase can be described for a limited time by a simple inverted harmonic oscillator model and that this limited time approaches infinity in the thermodynamic limit. Although we specifically discuss the Dicke model, the presented mechanism can be also used to describe dynamical quantum phase transitions in other systems and opens a new avenue in simulations of physical phenomena associated with an inverted harmonic oscillator.
\end{abstract}

\maketitle

\section{Introduction}
Quantum phase transitions describe abrupt changes of a system's properties while varying physical parameters, such as magnetization, at absolute zero temperature~\cite{sachdev2007quantum}. Depending on how fast the parameter is being changed, one can distinguish equilibrium and non-equilibrium quantum phase transitions and despite an almost dialectical relationship between these two types~\cite{henkel2008non}, the non-equilibrium one remains less well understood compared to the equilibrium one~\cite{dziarmaga2005PhysRevLett.95.245701,polkovnikov2005PhysRevB.72.161201,Review2011nonequlibrium,QKZM2019zoller}. The typical framework for studying non-equilibrium quantum phase transitions is a sudden change of some physical parameter in the Hamiltonian \cite{calabrese2006PhysRevLett.96.136801,sadler2006naturequenchBEC,collath2007PhysRevLett.98.180601,demarco2011quenchmott,chenaeu2012lightconespeed,polkovnikov2013quenchising,Klinder2015dickequench}. After such a quench, a system that was initially prepared in the ground state becomes a superposition of eigenstates of the quenched Hamiltonian which subsequently drives the evolution. In a complex system, such as the Dicke model, which consists of a collective spin coupled to a harmonic oscillator~\cite{garraway2011dicke}, one could expect a quench to lead to a complicated dynamical behavior---in certain cases, behavior with signatures of quantum chaos~\cite{brandes2003chaosdicke,chaos2011cejnar,2016hirschcahos,2019natcommundickerey}.

In this work, we show that for a limited and quantifiable time the dynamics after a quench can be mapped to dynamics of a simple inverted harmonic oscillator~\cite{jannussis1986harmonic,subramanyan2020physics}. Our model is based on an effective Hamiltonian that cannot be used to fully describe the equilibrium states~\cite{fehske2012PhysRevA.85.043821,plenio2015rabimodelqpt}. However, we show that it correctly describes dynamics for a limited time and under certain conditions. In fact, we show that in the thermodynamic limit {the non-equilibrium Dicke phase transition} becomes equivalent to a harmonic oscillator with a tunable frequency on both sides of the phase transition. By manipulating the parameters of the system and the spin polarization of the initial state, this tunable frequency can change from being purely real (normal phase), through being zero (critical point), to being purely imaginary (superradiant phase). Using state-of-the-art tools from quantum simulation~\cite{ruschhaupt2012PhysRevLett.109.100403,gietka2021PhysRevLett.126.160402}, we give a precise description of how to observe the physics of the inverted oscillator even far away from the thermodynamic limit. Moreover, our proposal opens a new avenue in simulations of physical phenomena associated with an inverted harmonic oscillator~\cite{subramanyan2020physics}. 

\begin{figure*}[htb!]
  \centering
\includegraphics[width=1\textwidth]{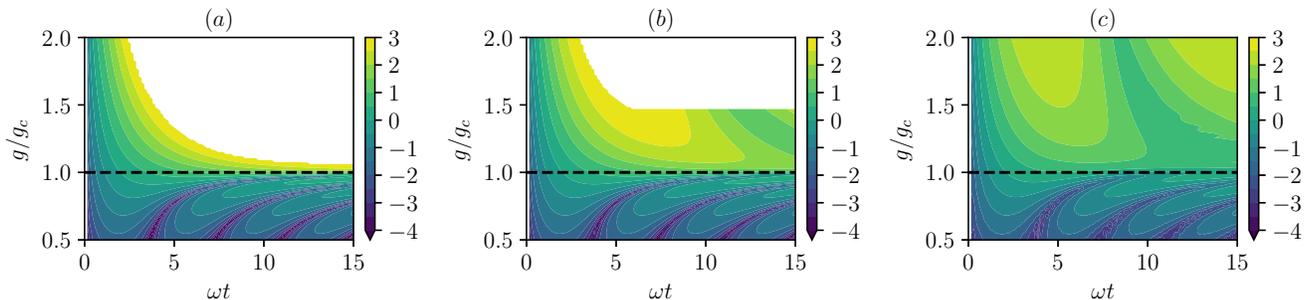}\hfill
\caption[fig1]{The logarithm of average number of photons as a function of $g/g_c$ and $t$ for $(a)$ $\sqrt{\Omega N/\omega} =100$, $(b)$ $\sqrt{\Omega N/\omega} \approx 31.6 $, and $(c)$  $\sqrt{\Omega N/\omega} =10$. The dashed-black line corresponds to the critical coupling in the thermodynamic limit. White regions correspond to unreliable numerical simulations after the boundary of the Hilbert space (set to $n = 3000$) has been reached. For simplicity, in the numerical simulations we set $N = 1$.}
\label{fig:fig1}
\end{figure*}


\section{Dicke model}
The Dicke model is a paradigmatic model describing the interaction of $N$ spin-$\frac{1}{2}$ systems (in general, two-level systems) with an energy splitting of $\Omega$ with a single mode field of frequency $\omega$~\cite{dicke1954PhysRev.93.99,HEPP1973dickesuperradiant,garraway2011dicke,brandes2012PhysRevLett.108.043003,2021dickedynamicsrey}. Its Hamiltonian can be written as (we set $\hbar$ to 1 throughout the entire manuscript)
\begin{equation}\label{eq:DM}
    \hat H = \omega \hat a^\dagger\hat a + \Omega\hat S_z + \frac{g}{\sqrt{N}}\left(\hat a +\hat a^\dagger\right)\hat S_x,
\end{equation}
where we have used the collective spin operators $\hat S_i = \sum_{n=1}^N \hat \sigma_i^{(n)}/2$ with $\hat \sigma_i^{(n)}$ being the $i$th Pauli matrix of the $n$th spin, and bosonic field creation and annihilation operators $\hat a^\dagger$ and $\hat a$, satisfying the bosonic commutation relations $[\hat a,\hat a^\dagger] = \hat{\mathds{1}}$. The parameter $g$ quantifies the collective interaction of $N$ spins with the single mode field. At a critical coupling strength $g=g_c\equiv \sqrt{\omega \Omega}$ in the limit of $\sqrt{\Omega N/\omega} \rightarrow \infty$ the Dicke model exhibits a quantum phase transition into a superradiant state. For $N \rightarrow \infty$ this limit becomes  thermodynamic ~\cite{garraway2011dicke}, which means that the quantum phase transition occurs for an arbitrary but finite ratio of $\Omega/\omega$. To gain more insight {into dynamics} one can apply the unitary transformation $\hat U = \exp\{i (g/\sqrt{N}\Omega)(\hat a+ \hat a^\dagger)\hat S_y\}$ to the Hamiltonian in Eq.~\eqref{eq:DM} and obtain an effective description~\cite{plenio2015rabimodelqpt,demler2021PhysRevLett.126.153603}
\begin{equation}\label{eq:effDM}
    \hat H_{\mathrm{eff}} \simeq  \omega \hat a^\dagger\hat a + \Omega\hat S_z + \frac{g^2}{2\Omega N }\left(\hat a +\hat a^\dagger\right)^2\hat S_z,
\end{equation}
which is exact in the limit of $\sqrt{\Omega N/\omega} \rightarrow \infty$. However, it is generally believed that this effective Hamiltonian is only applicable in the normal phase when $g<g_c$~\cite{plenio2015rabimodelqpt}. As the energy gap is given by $\omega\sqrt{1-g^2/g_c^2}$ the ground state energy of the effective Hamiltonian is imaginary for $g>g_c$ and a different effective description has to be employed in the superradiant phase. However, in this work we argue that, {following a quench from the groundstate in the normal phase to the superradiant phase} in the thermodynamic limit, the effective Hamiltonian from Eq.~(\ref{eq:effDM}) {can} also {be} the correct description in the superradiant phase as in the latter the ground state energy tends to $-\infty$ {(similarly as in the Holstein-Primakoff representation of the Dicke model)}. Moreover, even away from the thermodynamic limit ($\sqrt{\Omega N/\omega}\gg1$), one can use the Hamiltonian~(\ref{eq:effDM}) to describe the dynamics following a sudden quench to $g>g_c$ for times shorter than some critical time related to the initial state and energy of the ground state in the superradiant phase.

The initial state {(ground state for $g=0$)} consists of the vacuum state of the field $|0\rangle$ ($\hat a|0\rangle = 0$) and the collective spin down state, which is the eigenstate of the $\hat S_z$ operator with minimal eigenvalue, i.e., $\hat S_z |\psi\rangle = -\frac{N}{2}|\psi\rangle$. It can be easily checked that the transformation $\hat U = \exp\{i (g/\sqrt{N}\Omega)(\hat a+ \hat a^\dagger)\hat S_y\}$ leaves that state unchanged once $ \omega/\Omega \cdot g^2/g_c^2 \ll 1$ (see the Appendix~\ref{ap:inva} for a derivation). Therefore, since the effective Hamiltonian conserves the projection of the spin onto the $z$-axis, we can replace the operator with its lowest eigenvalue leading to
\begin{equation}\label{eq:effdmnospin}
    \hat H =  \left(\omega - \frac{g^2}{2\Omega }\right) \hat a^\dagger\hat a  - \frac{g^2}{4\Omega  }\left(\hat a^2 +\hat a^{\dagger2}\right).
\end{equation}
The above Hamiltonian describes a single-mode field with a modified frequency $\omega-g^2/2\Omega $ which is being squeezed with a strength equal to $g^2/2\Omega$. Both quantities are independent of $N$ and the two non-commuting mechanisms give rise to rich physics. We now consider a sudden quench from $g=0$ to some non-zero $g$. For $g$ below the critical value, the vacuum state is being squeezed and the number of photons increases~\cite{ZHANG199214squeezevacuum}. However, the additional rotation mechanism causes the state to unsqueeze after a $\pi/2$ rotation and brings it back to the vacuum with no photons (see the Appendix~\ref{ap:squeezing}). When $g$ approaches the critical value, the squeezing mechanism becomes dominant exactly at the critical point $g=g_c$ [note that at the critical point the first term in Hamiltonian~(\ref{eq:effdmnospin}) does not vanish] and the initial vacuum state starts to get squeezed more and more with further increasing $g$. This can be seen in Fig.~\ref{fig:fig1}, where we plot the average number of photons as a function of $g/g_c$ and $t$ for various ratios of $\sqrt{\Omega N/\omega}$ and for an initial vacuum state evolved with the Hamiltonian from Eq.~(\ref{eq:DM}). Looking at Fig.~\ref{fig:fig1}, we can identify parameter regions of squeezing-rotation (oscillations of the number of photons) below the critical point and for the case of $\sqrt{\Omega N/\omega}$ very large [see panel $(a)$] continuous squeezing (increase of the number of photons) above the critical point. Note that white regions in this plot correspond to unreliable numerical simulations after the boundary of the Hilbert space (set to $n = 3000$) has been reached due to the large increase in photon production.
However,  going away from the limit $\sqrt{\Omega N/\omega} \rightarrow \infty$ [see panels $(b)$ and $(c)$], the number of photons does not continuously increase in the superradiant regime, which puts a limit on the time for which the effective Hamiltonian from Eq.~(\ref{eq:effDM}) 
can be used to describe the dynamics governed by Eq.~(\ref{eq:DM}) for $g>g_c$. The limiting time is heralded when the growth of the photon number slows down. However, for a limited time the Hamiltonian from Eq.~(\ref{eq:effDM}) gives rise to the correct qualitative behavior even if $g>g_c$. Before discussing the time limitation for that description let us now have a look at the system from another perspective.
\begin{figure*}[htb!]
  \centering
\includegraphics[width=1\textwidth]{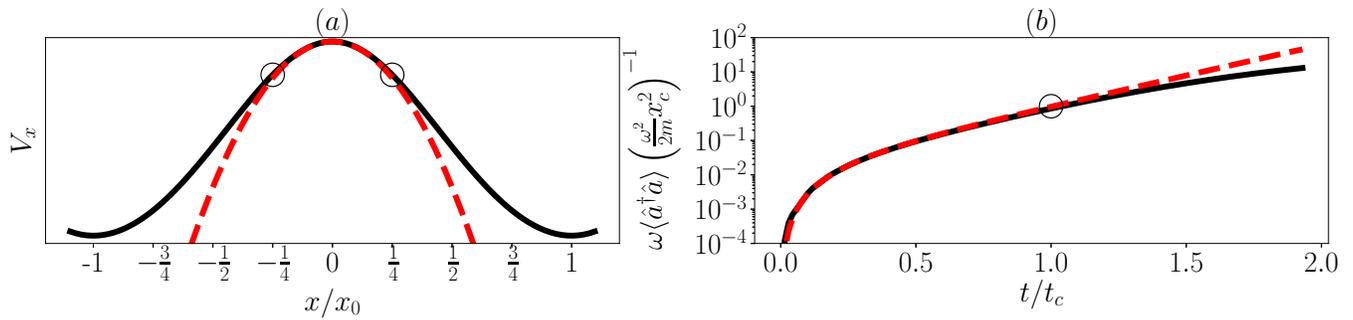}\hfill
\caption[fig2s]{Phenomenological condition for simulating the inverted oscillator with the Dicke model. $(a)$ shows the effective double-well potential (full black line) overlaid on an inverted oscillator potential (dashed red line). The black circles indicate the critical position until which both models agree. $(b)$ The critical position can be translated to a condition on the number of photons which can be then related to a critical time for which the Dicke model can be used as a simulator. This time is indicated by the black circle.}
\label{fig:fig2s}
\end{figure*}


\section{Inverted harmonic oscillator}
Additional insight can be gained by rewriting the Hamiltonian~(\ref{eq:effdmnospin}) in terms of the pseudo-position and pseudo-momentum operators (quadratures), $\hat X = (\hat a + \hat a^\dagger)/\sqrt{2}$ and $\hat P = (\hat a - \hat a^\dagger)/\sqrt{2}i$, as
\begin{equation}
    \hat H = \frac{\omega}{2}\hat P^2 +\frac{1}{2\omega}\left({\omega^2}-\frac{\omega g^2}{\Omega}\right)\hat X^2.
\end{equation}
This shows that the system dynamics behaves as a particle with mass $1/\omega$ in a harmonic oscillator potential of frequency $\omega\sqrt{1-g^2/g_c^2}$. If $g<g_c$ the frequency is real and the energy gap above the ground state is proportional to $\omega\sqrt{1-g^2/g_c^2}$. If $g>g_c$ the frequency becomes purely imaginary and the energy gap cannot be defined as there is no ground state of the Hamiltonian. In other words, {for a quench from the $g=0$ groundstate}, the Dicke model in the thermodynamic limit becomes an inverted harmonic oscillator in the superradiant phase. With this picture in mind, we can easily explain the dynamics seen in the previous section. As the coupling parameter is increased towards the critical value the frequency of the oscillations (squeezing-rotation) decreases, which corresponds to the harmonic potential being flattened. At the critical point $g = g_c$ the frequency of the oscillations becomes zero as the harmonic potential becomes completely flat (free particle). As the coupling becomes greater than the critical coupling the frequency becomes purely imaginary, or in other words, the harmonic potential flips upside-down giving rise to an unbounded fall (squeezing). However, away from the thermodynamic limit, i.e., when $\sqrt{\Omega N/\omega} \gg 1$, this picture does only apply for finite periods, and we will attempt to clarify the limitations in the following.


\begin{figure*}[htb!]
  \centering
\includegraphics[width=1\textwidth]{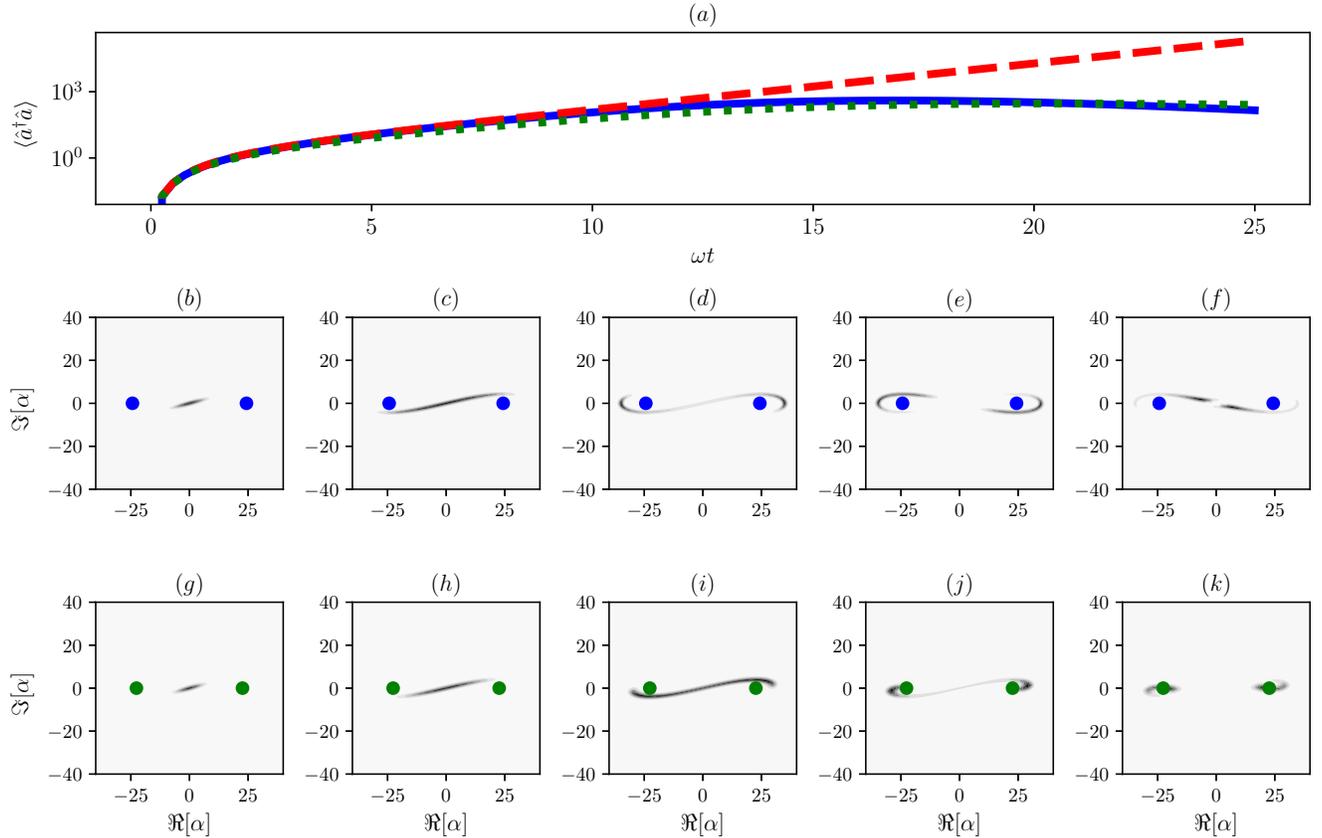}\hfill
\caption[fig0]{$(a)$ Comparsison between an inverted harmonic oscillator (dashed-red), the isolated (solid-blue), and open (dotted-green) Dicke model. The Husimi function of the field for five different values of $\omega t =\{5,10,15,20,25\}$ for isolated dynamics is depicted in $(b)$-$(f)$ and for open dynamics is depicted in $(g)$-$(k)$. The blue and green dots indicate the bottom of the effective potential for the isolated and open case, respectively. In the simulation we have set $\sqrt{\Omega N/\omega} = 100$, $\kappa = 0.1 \omega$, $\gamma = 0.01\Omega$, $g = 1.03g_c$ and $N = 1$.}
\label{fig:fig0}
\end{figure*}

\section{The connector operator}
As discussed in the introduction, we will use the tools from quantum simulation theory~\cite{ruschhaupt2012PhysRevLett.109.100403,gietka2021PhysRevLett.126.160402} to get a deeper understanding of how applicable the inverted harmonic oscillator picture is. The condition for a quantum simulator can be expressed as
\begin{align}\label{eq:condgen}
	   \langle \psi| e^{i  \hat H_{\mathrm{QS}}t} e^{-i  \hat H_{\mathrm{T}}t} | \psi \rangle = \langle \psi| e^{i  \hat h(t)} | \psi \rangle = e^{i  \xi(t)},
\end{align}
where $\hat H_{\mathrm{QS}}$ and $\hat H_{\mathrm{T}}$ are the simulator and target Hamiltonian, respectively, $\xi(t)$ is a real-valued function of time, and $\hat h(t)$ is the connector operator which can be expressed using the Baker-Campbell-Hausdorff formula as
\begin{align}\label{eq:BCH}
	\hat h(t) = t\left(\hat H_{\mathrm{QS}} -\hat H_{\mathrm{T}} \right)+\frac{i t^2}{2}\left[\hat H_{\mathrm{QS}}, -\hat H_{\mathrm{T}}   \right]+\ldots,
\end{align}
where $\left[\bullet,\bullet \right]$ stands for the commutator and $\ldots$ indicates terms involving higher order commutators of $\hat H_{\mathrm{QS}}$ and $\hat H_{\mathrm{T}}$. In general, for states $|\psi\rangle$ which are eigenstates of $h(t)$ but not eigenstates of $\hat H_{\mathrm{QS}}$ and $\hat H_{\mathrm{T}}$, the two Hamiltonians will realize the same dynamics. For the Dicke model~(\ref{eq:DM}) and the effective description from Eq.~(\ref{eq:effDM}), the connector operator becomes exactly 0 in the thermodynamic limit. However, for a limited time the connector operator also converges to 0 indicating the transitional validity of the Hamiltonian \eqref{eq:effDM}.

The condition for the time during which the Hamiltonian \eqref{eq:effDM} is valid can be expressed as a relation between the instantaneous average number of photons and the number of photons in the ground state of the Dicke model. The latter can be calculated by rewriting the Hamiltonian using the pseudo-position and pseudo-momentum operators
\begin{equation}
    \hat H = \frac{\omega}{2} \hat p^2 + \frac{\omega}{2} \hat x^2 + \Omega \hat S_z + \frac{\sqrt{2} g}{\sqrt{N}}\hat x \hat S_x.
\end{equation}
The part containing the spin operators can be easily diagonlized, and its ground state energy can be found to be
\begin{equation}
   \left(\Omega \hat S_z + \frac{\sqrt{2} g}{\sqrt{N}}\hat x \hat S_x \right)|\sigma\rangle = \left( -\frac{1}{2} \sqrt{N} \sqrt{2 g^2 \hat x^2+N \Omega ^2}\right)|\sigma\rangle ,
\end{equation}
where $|\sigma\rangle$ is the spin ground state. Projecting the Hamiltonian onto the spin ground state yields 
\begin{equation}
    \hat H = \frac{\omega}{2} \hat p^2 + \frac{\omega}{2} \hat x^2 -\frac{1}{2} \sqrt{N} \sqrt{2 g^2 \hat x^2+N \Omega ^2}.
\end{equation}
The above Hamiltonian describes a particle with mass $1/\omega$ moving in a double-well potential given by $\hat V(x) = \frac{\omega}{2} \hat x^2 -\frac{1}{2} \sqrt{N} \sqrt{2 g^2 \hat x^2+N \Omega ^2}$. The minima of such an effective potential can be evaluated to be
\begin{equation}
    x_0 = \pm \sqrt{\frac{N \Omega }{2 \omega }}\sqrt{\frac{g^2}{g_c^2}-\frac{g_c^2}{g^2}}.
\end{equation}
From a simple phenomenological argument (see Fig.~\ref{fig:fig2s}) the critical position for the observation of the inverted harmonic oscillator will be a quarter of the minima position, i.e., $x_c = x_0/4$. 

According to this position we can establish an energy relation which has to be satisfied 
\begin{equation}
    \frac{\omega}{2}\langle \hat p^2 + \hat x^2  \rangle < \frac{\omega}{2}\hat x_c^2 = \frac{\omega}{32}\hat x_0^2,
\end{equation}
which upon a simplification yields
\begin{equation}
    \left\langle \hat a^\dagger \hat a \right\rangle < \frac{1}{32} {\frac{N \Omega }{ \omega }}\left({\frac{g^2}{g_c^2}-\frac{g_c^2}{g^2}}\right).
\end{equation}
For a special case of $g = \sqrt{2}g_c$ and initial state being the vacuum state, we can simplify the condition to
$
    \sinh^2(\omega t) <  {3}/{64}\cdot {{N \Omega }/{ \omega }},
$
which for $\omega t>1$ becomes
\begin{equation}
     t < t_c \equiv  \ln\left[\sqrt{\frac{3}{16}} \sqrt{{\frac{N \Omega }{ \omega }}}\right]\omega^{-1}.
\end{equation}

The above time limitations can be intuitively understood as a time until which the time-evolved state \emph{feels} the bottom of the effective double-well potential after a quench out of the equilibrium position at $x=0$ (see Fig.~\ref{fig:fig2s}). In the thermodynamic limit, the overlap increases as the minima of the double-well potential are located at plus and minus infinity which means that the double-well potential becomes in fact identical to the inverted harmonic oscillator (see Appendix~\ref{ap:connector}).

The comparison between the dynamics of the Dicke model and the inverted oscillator is presented in Fig.~\ref{fig:fig0} where we show the number of photons as a function of time $(a)$ and squeezing of photons as a function of time for isolated $(b)$-$(f)$ and open $(g)$-$(k)$ dynamics through the Husimi function. For the case of isolated dynamics the panels $(b)$ and $(c)$ depict squeezing (exponential growth of the photon number) due to the inverted harmonic oscillator potential and its termination as the state starts to feel the potential minima (blue dots). The slowdown of squeezing and reaching of the potential minima can be seen in panels $(d)$ and $(e)$. Finally, panel $(f)$ depicts anti squeezing (decrease of the number of photons). For the case of open dynamics (see the Appendix~\ref{ap:deco}), we observe initially the similar behavior in panels $(g)$ to $(i)$, but with a slower  growth of the photon number. Furthermore, as the system dissipates the energy, the state \emph{cannot climb back} to the local maximum located at $x=0$ and eventually becomes an incoherent mixture of two coherent states with opposite amplitudes (symmetry breaking) which can be seen in panels $(j)$ and $(k)$.


\section{Conclusions \& Outlook}
We have shown that the dynamics following a quench {from the groundstate in the normal phase ($g=0$)} into the superradiant phase in the Dicke model can be described by a model of an inverted harmonic oscillator and argued that in the thermodynamic limit the Hamiltonian of the system becomes equivalent to the inverted harmonic oscillator. The test of results can be readily performed in quantum simulators which can realize the Dicke model \cite{domokos2002PhysRevLett.89.253003,vuletic2003PhysRevLett.91.203001,baumann2010dicke,Mottl1570,PhysRevA.97.043858,PhysRevA.97.042317,engels2014dickesocbec,Klinder2015dickequench}, and in quantum simulators which realize the quantum Rabi model ($N=1$) \cite{solano2019strongcouplingreview,kockum2019ultrastrong} using various platforms including cold atoms \cite{rauschnbeutel2018observationofultrastrong}, trapped ions \cite{kihwan2018qrmtrappedion}, superconducting circuits \cite{braumuller2017analog}, and electrons trapped on liquid Helium~\cite{konstantinov2019PhysRevLett.122.176802}. The presented mechanism should also apply to an arbitrary quantum system exhibiting a quantum phase transition associated with an effective double-well or \emph{sombrero} potential.

The presented idea may be used as a new tool to study quantum phase transitions \cite{zurekzoller2005qptdynamics,sachdev2007quantum}, the quantum Kibble-Zurek mechanism~\cite{Kibble_1976,KIBBLE1980183,zurek1985,ZUREK1996177,sondhi2012kZPhysRevB.86.064304} both theoretically and experimentally, and the quantum Lieb–Robinson bound for how fast correlations can spread in a quantum system~\cite{chenaeu2012lightconespeed}. The results of this work can be also applied in quantum metrology for precise measurements of $\omega$ and $\Omega$~\cite{PhysRevLett.124.120504_2020_garbeqcm,PhysRevLett.2021_dyanmicqm,tobe} as well as for the preparation of squeezed states, simulations of the inflation of the early Universe \cite{PhysRevD.50.4807,PhysRevD.42.3413,PhysRevD.32.1899,PhysRevX.8.021021_2018_rapidbecexp}, the Unruh effect~\cite{PhysRevD.14.870,RevModPhys.80.787}, Hawking radiation~\cite{hawkingrad}, quantum chaos~\cite{watanabe2020,yan2021chaos}, and potentially to study many other aspects of physics in which the inverted harmonic oscillator has been harnessed as an underlying mechanism (see Ref.~\cite{subramanyan2020physics} for the recent review). 

Finally, we would like to point out that even though this model bears a resemblance to the Landau theory of phase transition~\cite{landau1937theory}, the presented description is fully quantum.


\begin{acknowledgments}
Simulations were performed using the open-source QuantumOptics.jl framework in Julia~\cite{kramer2018quantumoptics}. K.G. would like to acknowledge discussions with Friederike Metz, Ayaka Usui, Lewis Ruks, Farokh Mivehvar, and Micha{\l} B{\c a}czyk. This work was supported by the Okinawa Institute of Science and Technology Graduate University. K.G. acknowledges support from the Japanese Society for the Promotion of Science (P19792).
\end{acknowledgments}

\onecolumngrid
\appendix
\section{Invariance of the initial state under the transformation}\label{ap:inva}
In order to clearly see the effect of the inverted oscillator, the initial state has to remain unaltered under the transformation
\begin{equation}
    \hat U = \exp\left(i \frac{g}{\sqrt{N}\Omega}\left(\hat a+ \hat a^\dagger\right)\hat S_y\right).
\end{equation}
The condition for this to happen can be found by considering the overlap
\begin{equation}
    \langle \downarrow \!0|\exp\left(i \frac{g}{\sqrt{N}\Omega}\left(\hat a+ \hat a^\dagger\right)\hat S_y\right)| 0\!\downarrow\rangle,
\end{equation}
where $| 0\!\downarrow\rangle \equiv | 0\rangle\otimes|\!\downarrow\rangle$ is the ground state of the system for $g=0$. Expanding the exponential function to fourth order (the higher order terms can be safely neglected because of $1/N$ dependence) we get 
\begin{equation}
\begin{split}
\langle \downarrow \!0|&\exp\left(i \frac{g}{\sqrt{N}\Omega}\left(\hat a+ \hat a^\dagger\right)\hat S_y\right)| 0\!\downarrow\rangle = 
    \langle \downarrow \!0| 0\!\downarrow\rangle +  \langle \downarrow \!0|i \frac{g}{\sqrt{N}\Omega}\left(\hat a+ \hat a^\dagger\right)\hat S_y| 0\!\downarrow\rangle  + \langle \downarrow \!0|i^2 \frac{g^2}{N\Omega^2}\left(\hat a+ \hat a^\dagger\right)^2\hat S_y^2| 0\!\downarrow\rangle \\
   & + \langle \downarrow \!0|i^3 \frac{g^3}{\sqrt{N^3}\Omega^3}\left(\hat a+ \hat a^\dagger\right)^3\hat S_y^3| 0\!\downarrow\rangle + \langle \downarrow \!0|i^4 \frac{g^4}{N^2\Omega^4}\left(\hat a+ \hat a^\dagger\right)^4\hat S_y^4| 0\!\downarrow\rangle + \mathcal{O}\left(\frac{g^5}{\sqrt{N^5}\Omega^5}\right).
\end{split}
\end{equation}
Since terms with odd powers yield zero after the evaluation, we can simplify the above overlap to
\begin{equation}
\begin{split}
   \langle \downarrow \!0|\exp\left(i \frac{g}{\sqrt{N}\Omega}\left(\hat a+ \hat a^\dagger\right)\hat S_y\right)| 0\!\downarrow\rangle \approx 1 - \frac{g^2}{4 \Omega^2} +  \frac{g^4}{g_c^4} \frac{\omega^2}{\Omega^2}\left(\frac{9}{16} - \frac{24}{N}\right).
   \end{split}
\end{equation}
Therefore if we want the state $| 0\!\downarrow\rangle$ to remain invariant, the following condition has to be fulfilled
\begin{equation}
  \frac{1}{4} \frac{g^2 }{g_c^2 }\frac{ \omega}{\Omega}\ll 1.
\end{equation}
If however, the initial state is not invariant under the transformation, the system will still simulate the inverted harmonic oscillator but for a transformed state, {which however seems impractical. This can be seen explicitly if we transform the Schr\"odinger equation
\begin{equation}
i\partial_t\hat U | \psi \rangle = \hat H_{\mathrm{eff}}\hat U |\psi \rangle \rightarrow i \partial_t |\tilde{\psi}\rangle =  \hat H_{\mathrm{eff}} |\tilde{\psi} \rangle,
\end{equation}
where $|\tilde{\psi} \rangle$ is the transformed state $\hat U | \psi \rangle$.}

\section{The connector operator}\label{ap:connector}
\begin{figure*}[htb!]
  \centering
\includegraphics[width=1\textwidth]{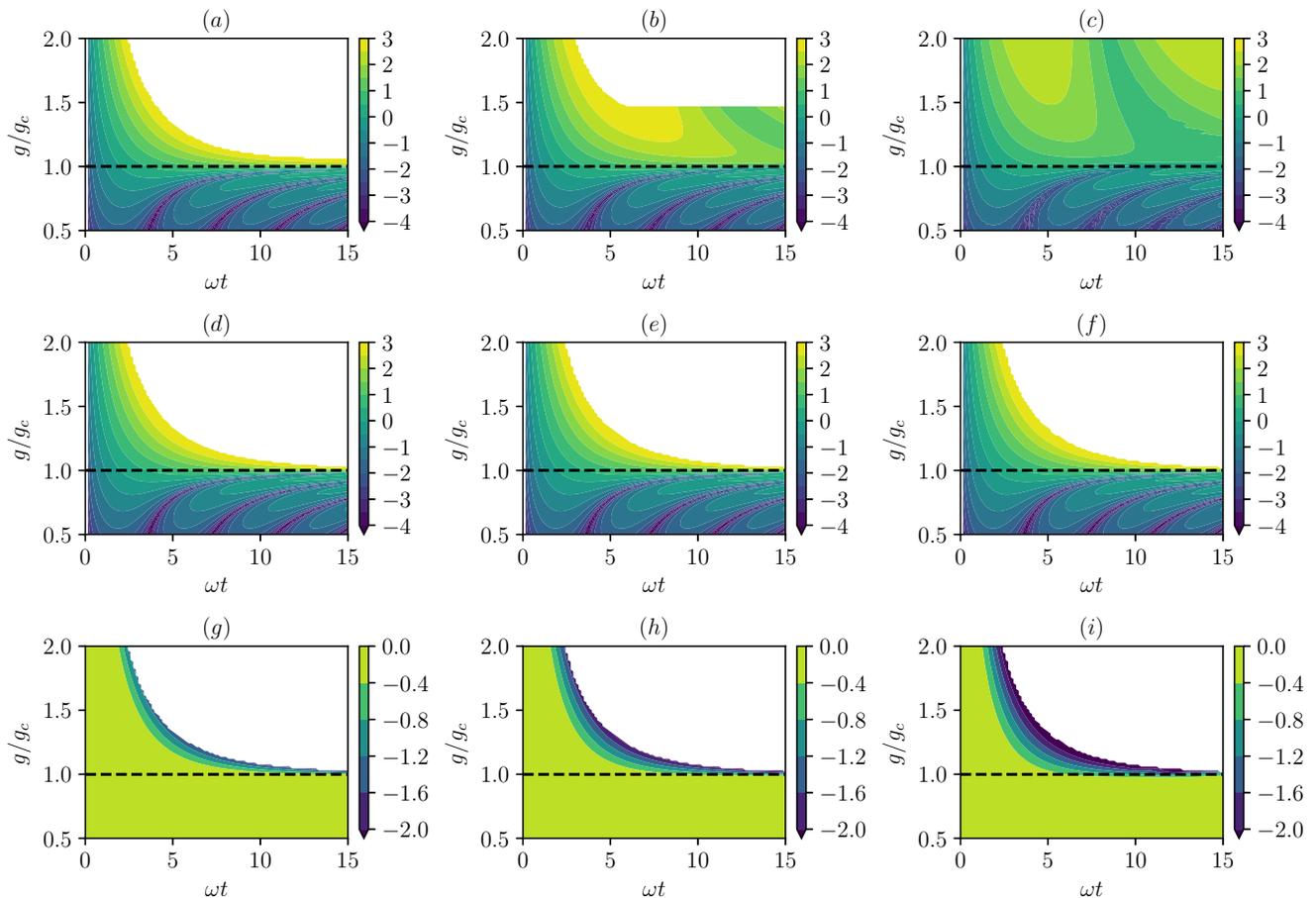}\hfill
\caption[fig1s]{Extension of Fig.~1 from the main text. The Dicke model $(a)$-$(c)$ can be used to simulate the physics of the inverted harmonic oscillator $(d)$-$(f)$. The logarithm of fidelity between the states generated by these two Hamiltonians $(g)$-$(i)$ sets a time limitation for the simulator to work properly. White regions correspond to unreliable numerical simulations after the boundary of the Hilbert space (set to $n= 3000$) has been reached.}
\label{fig:fig1s}
\end{figure*}
The condition for simulating the inverted oscillator with the Dicke model can be expressed with the use of the fidelity
\begin{equation}
    \langle \psi |  e^{i  \hat H_{\mathrm{DM}}t} e^{-i  \hat H_{\mathrm{IO}}t} |\psi \rangle = e^{i \xi(t)},
\end{equation}
where the subscripts DM and IO correspond to the Dicke model and the inverted oscillator, respectively. As discussed in the main text, if the imaginary part of $\xi(t)$ is negligible, one system can simulate the physics of the other one even if the Hamiltonians are different. In Fig.~\ref{fig:fig1s}, we plot the logarithm of $|e^{i \xi(t)}|$. When this logarithm is equal to 0, it means that the initial state is an eigenstate of the connector operator
\begin{align}\label{eq:connectorfull}
	\hat h(t) = t\left(\hat H_{\mathrm{DM}} -\hat H_{\mathrm{IO}} \right)+\frac{i t^2}{2}\left[\hat H_{\mathrm{DM}}, -\hat H_{\mathrm{IO}}   \right]+\ldots,
\end{align}
with eigenvalue equal to multiples of $\pi$ including 0. As can be seen in Fig.~\ref{fig:fig1s}, by increasing the $\sqrt{N \Omega/\omega}$, we are increasing the critical time for which the Dicke model can simulate the inverted harmonic oscillator. In the limit of $\sqrt{N \Omega/\omega} \rightarrow \infty$, the Dicke model can realize exactly the same dynamics as the inverted harmonic oscillator. In other words, by increasing $\sqrt{N \Omega/\omega}$ the initial state can remain an eigenstate of the connector operator for longer times as the connector operator becomes an identity in the thermodynamic limit.

\section{Squeezing oscillations in the normal phase}\label{ap:squeezing}
\begin{figure*}[htb!]
  \centering
\includegraphics[width=1\textwidth]{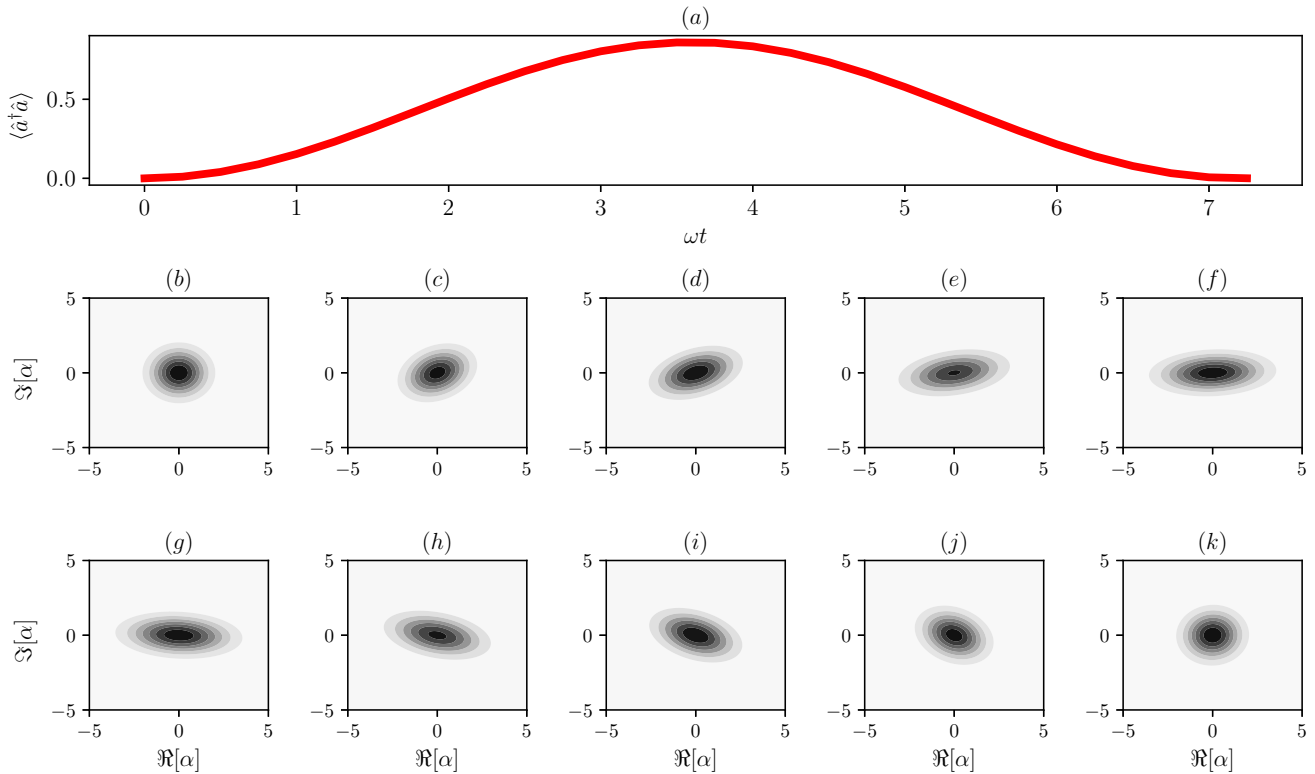}\hfill
\caption[fig4osc]{Squeezing (photon number) oscillations after the quench to the normal phase. Cf.~Fig.~\ref{fig:fig2}. $(a)$ depicts the number of photons as a function of time.  The Husimi function of the field for ten different values of $\omega t =\{0,0.8,1.6,2.4,3.2,4.0,4.8,5.6,6.4,7.2\}$ for isolated dynamics is depicted in $(b)$-$(k)$ and shows the squeezing oscillations. In the simulation we have set $\sqrt{\Omega N/\omega} = 100$, $g = 0.9g_c$, and $N = 1$.}
\label{fig:fig4osc}
\end{figure*}

In the main text, we write that the Dicke model after the quench to the normal phase exhibits squeezing oscillations or oscillations of the number of photons if the initial state is the field vacuum state. In order to understand it, one has to realize that squeezing a vacuum leads to a state with a non-zero number of photons. Therefore squeezing and anti-squeezing (squeezing oscillations) will lead to oscillations of the number of photons. This can be seen in Fig.~\ref{fig:fig4osc} where we show the number of photons as a function of time $(a)$, and the Husimi function for ten different times depicting a period of squeezing oscillations $(b)$-$(k)$.
\section{Effect of decoherence} \label{ap:deco}
\begin{figure*}[htb!]
  \centering
\includegraphics[width=1\textwidth]{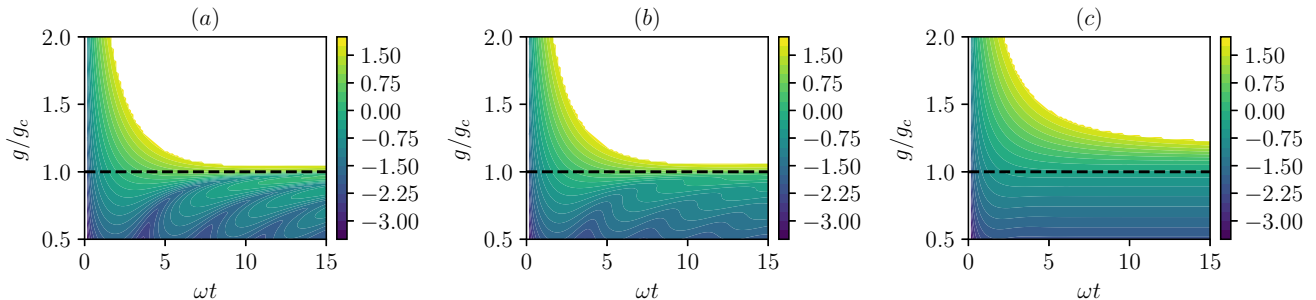}\hfill
\caption[fig2]{The logarithm of the average number of photons as a function of $g/g_c$ and $t$ for $(a)$ $\kappa = 0.01\omega$ and $ \gamma = 0.1\Omega$, $(b)$ $\kappa = 0.1\omega$ and $ \gamma = 0.1\Omega$, and $(c)$ $\kappa = 0.5\omega$ and $ \gamma = 0.1\Omega$.  The dashed-black line corresponds to the critical coupling in the thermodynamic limit. White regions correspond to unreliable numerical simulations after the boundary of the Hilbert space (set to $n = 200$) has been reached. In the numerical simulations we set $\sqrt{\Omega N/\omega} \approx 31.6$ and $N = 1$.}
\label{fig:fig2}
\end{figure*}

The discussion from the main text revolves around isolated systems. However, from an experimental perspective, one has to include decoherence which is a consequence of an inability to perfectly isolate a quantum system from its environment. In order to account for these typically unwanted effects, we use the Lindblad Master equation approach
\begin{equation}\label{eq:LME}
    \frac{\mathrm{d} \hat \rho}{\mathrm{d}t} = - i [\hat H, \hat \rho ] +2\kappa\left( \hat a \hat \rho \hat a^\dagger - \{\hat a^\dagger \hat a,\hat \rho\}\right) +2 \gamma \left( \hat S_- \hat \rho \hat S_+ - \{\hat S_+ \hat S_-,\hat \rho\}\right),
\end{equation}
where $\kappa$ and $\gamma$ account for the damping of mode $\hat a$ and damping of the spin, respectively, with $\hat S_{-} =\hat S_+^\dagger = \hat S _x - i \hat S_y$ being the spin lowering operator. The effect of spin damping should be negligible as the initial state is the lowest energy spin eigenstate and for a limited time the system conserves the projection of the spin onto the $z$-axis. If this was not the case, the damping mechanism would bring the state of the spin into its lowest energy eigenstate and increase the squeezing rate (change the frequency of the inverted oscillator). In both phases the photon damping will cause the system to eventually reach a steady state which is not a vacuum state as the system is driven by a non-zero $g$. The results of the numerical simulations for both phases are presented in Fig.~\ref{fig:fig2}, where we plot the average number of photons as a function of $g/g_c$ and $t$ for a fixed ratio $\sqrt{\Omega N/\omega}$ for an initial vacuum state evolved with Master equation from Eq.~(\ref{eq:LME}) using the Dicke Hamiltonian~(1) with three different values of $\kappa$. As expected, by introducing and increasing the photon loss rate $\kappa$ the number of produced photons is reduced and eventually reaches a steady state [see Fig.~\ref{fig:fig2}$(c)$]. Squeezing of photons as a function of time including the inverted harmonic oscillator phase and reaching the potential minima phase (associated with a steady state) is depicted in Fig.~3$(g)$-$(k)$. 

\twocolumngrid

\end{document}